\documentclass[12pt]{iopart}
\usepackage{iopams}
\usepackage{amssymb}
\usepackage{graphicx}
\usepackage{portland}
                       
\def\beq{\begin{equation}}
\def\eeq{\end{equation}}
\def\bea{\begin{eqnarray}}
\def\eea{\end{eqnarray}}
\def\mr{\mathrm}

\def\ug{\,=\,}
\def\lp{\left(}
\def\rp{\right)}
\def\lP{\left[}
\def\rP{\right]}
\def\pr{\prime}

\def\ltord{\hbox{$\;\raise.4ex\hbox{$<$}\kern-.75em\lower.7ex\hbox{$\sim$}
                       \;$}}
\def\gtord{\hbox{$\;\raise.4ex\hbox{$>$}\kern-.75em\lower.7ex\hbox{$\sim$}
                       \;$}}

\begin{document}
\title[Primordial black hole formation in the radiative era]
{Primordial black hole formation in the radiative era: investigation 
of the critical nature of the collapse}
\author{Ilia Musco${}^{1,2}$, John C. Miller${}^{2,3}$ and 
Alexander G. Polnarev${}^{4}$}

\address{${}^1$ Centre of Mathematics for Applcations, Department of
Mathematics, University of Oslo, PO Box 1053 Blindern, NO-0316 Oslo, 
Norway\\
${}^2$ SISSA, International School for Advanced Studies and INFN, 
Via Beirut 2-4, I-34014 Trieste, Italy\\
${}^3$ Department of Physics (Astrophysics), University of Oxford, Keble
Road, Oxford OX1 3RH, UK\\
${}^4$ Astronomy Unit, Queen Mary University of London, Mile End Road, 
London E1 4NS, UK\\}

\begin{abstract}
 Following on after two previous papers discussing the formation of 
primordial black holes in the early universe, we present here results 
from an in-depth investigation of the extent to which primordial black 
hole formation in the radiative era can be considered as an example of 
the critical collapse phenomenon. We focus on initial 
supra-horizon-scale perturbations of a type which could have come from 
inflation, with only a growing component and no decaying component. In 
order to study perturbations with amplitudes extremely close to the 
supposed critical limit, we have modified our previous computer code 
with the introduction of an adaptive mesh refinement scheme. This has 
allowed us to follow black hole formation from perturbations whose 
amplitudes are up to eight orders of magnitude closer to the threshold 
than we could do before. We find that scaling-law behaviour continues 
down to the smallest black hole masses that we are able to follow and 
we see no evidence of shock production such as has been reported in 
some previous studies and which led there to a breaking of the 
scaling-law behaviour at small black-hole masses. We attribute this 
difference to the different initial conditions used. In addition to 
the scaling law, we also present other features of the results which 
are characteristic of critical collapse in this context.
 \end{abstract}


\submitto{\CQG}

\maketitle

%=======================================================
\section{Introduction}
 It is possible that a population of black holes may have been formed 
during the radiative era of the early universe by means of gravitational 
collapse of cosmological perturbations. This was first proposed by 
Zel'dovich \& Novikov (1969) \cite{Zeldovich} and then by Hawking (1971) 
\cite{Hawking}. (Note that throughout this paper, we use the word 
``perturbation'' to refer to generic deviations away from the uniform 
background solution for the universe. These deviations are only being 
considered as small or linear where that is explicitly stated.) The idea 
of these primordial black holes (PBHs) has subsequently attracted 
continuing attention, even if they have sometimes been regarded as 
uninteresting on the grounds that perturbations with amplitudes 
sufficiently large to form them are likely to have been extremely rare. 
Following the initial proposals, the possible formation process was then 
investigated further by Carr \cite{Carr1, Carr2} using an ingenious 
simplified model in which he considered an overdense spherical region, 
described as part of a closed Friedmann-Robertson-Walker (FRW) universe, 
surrounded by a spatially-flat FRW expanding background. Results from this 
simplified model suggested that PBH formation would occur if the 
perturbation amplitude $\delta$, defined as the relative mass excess 
inside the overdense region measured when it had the same scale as the 
cosmological horizon, is greater than a threshold value of about 1/3. A 
more precise determination of this threshold value, usually referred to as 
$\delta_c$, requires a numerical computation and this was first done by 
Nadezhin, Novikov \& Polnarev (1978) \cite{Nadezhin} and by Bicknell \& 
Henriksen (1979) \cite{Bicknell}. In the 1980s, attention shifted mainly 
to the cosmological consequences that a population of PBHs would have, and 
this was extensively studied in various scenarios (see, for example, the 
recent review by Carr \cite{Carr3}). In 1999 Niemeyer \& Jedamzik 
\cite{Jedamzik1} returned to the issue of making numerical calculations of 
the PBH formation process, and they were subsequently followed in this by 
other groups \cite{Shibata, Hawke} including ourselves \cite{Musco1, 
Musco2}. All of these studies confirmed the overall picture of PBH 
formation, showing that the precise value of $\delta_c$ depends on the 
initial perturbation profile.

In practice, the main perturbations of interest here are ones coming from 
inflation which then re-enter the cosmological horizon at later times. For 
understanding the cosmological impact of PBHs, it is necessary to 
determine the probabilities of different types of perturbation profile 
(see Hidalgo \& Polnarev \cite{Hidalgo}, and references therein) and the 
corresponding values of $\delta_c$. However it is also necessary to have a 
consistent understanding of the collapse process and of the dependence of 
the PBH mass on the perturbation amplitude. An important contribution to 
this was made by Niemeyer \& Jedamzik \cite{Jedamzik1, Jedamzik2} who 
showed that the masses of PBHs produced in the radiative era by 
perturbations with a given profile type, follow the scaling-law behaviour 
of critical collapse, first discovered for idealized conditions by 
Choptuik \cite{Choptuik}, i.e. the masses of the black holes produced 
follow a power law $M_{BH} \propto (\delta - \delta_c)^\gamma$ if $\delta$ 
is close enough to $\delta_c$. If one considers a perfect fluid with an 
equation of state of the form $p = we$ (where $p$ is the pressure and $e$ 
is the energy density), then the value of the critical exponent $\gamma$ 
depends on the value of $w$ \cite{Choptuik2}. Niemeyer \& Jedamzik found a 
scaling law with $\gamma \simeq 0.36$ for the models which they studied, 
agreeing with the value of $\gamma$ already found for idealized critical 
collapse in a radiation fluid (having $w=1/3$) \cite{Evans}. However, 
their code was only able to handle cases with a fairly small range of 
$\delta - \delta_c$ (and hence of PBH masses), leaving open the question 
of whether the scaling law would continue down to very low values of 
$M_{BH}$. In 2002 this issue was taken up by Hawke \& Stewart \cite{Hawke} 
who used a sophisticated purpose-built code which was able to handle a 
much wider range of values of $\delta - \delta_c$ and deal with the strong 
shocks that they found arising for cases near to the critical limit 
($\delta \to \delta_c$). They too found a scaling law over a certain range 
of PBH masses (spanning a factor of $\sim 100$) but with deviations away 
from this for higher and lower masses. In particular, for the 
perturbations having amplitudes closest to the critical limit, they found 
that the relationship levelled off at a minimum mass of $\sim 10^{-3.5}$ 
of the cosmological horizon mass at their initial time.

An issue with all of these calculations concerns the specification of 
initial conditions. Niemeyer \& Jedamzik used non-linear horizon-scale 
perturbations as their initial conditions, while Hawke \& Stewart used 
non-linear perturbations at a sub-horizon-scale. While completely 
self-consistent, these were importantly different from the type of 
perturbation coming from inflation in that, as pointed out by Shibata \& 
Sasaki \cite{Shibata}, they included a decaying component which would have 
been absent in perturbations coming from inflation. Bearing this in mind, 
we ourselves then made calculations in 2005 (\cite{Musco1}, hereafter 
referred to as Paper 1) for which the initial conditions were linear 
supra-horizon-scale perturbations of the energy density. During the 
subsequent evolution, the decaying component of the perturbation had time 
to become negligible before the perturbation passed inside the 
cosmological horizon (at the ``horizon-crossing'' time) and started its 
non-linear evolution, making the treatment of the eventual black-hole 
formation a better representation of what would happen for perturbations 
coming from inflation. (With this, we again demonstrated a scaling-law, 
over a rather small range, but found different values of $\delta_c$ from 
before.) However, the best thing to do is clearly to specify from the 
beginning perturbations containing only a growing component. We addressed 
this in a subsequent paper in 2007 \cite{Musco2}, where the initial 
conditions were again specified as linear supra-horizon-scale 
perturbations of the fluid parameters, but now using an asymptotic 
quasi-homogeneous solution \cite{Lifshitz}, which gives a consistent pure 
growing behaviour. When the perturbation length-scale is much larger than 
the cosmological horizon, the curvature perturbation commonly denoted by 
$\zeta$ can be non-linear (as necessary in order to subsequently produce a 
black hole) while the perturbations of the hydrodynamical quantities are 
still in the linear regime. In this situation, the coupled system of 
Einstein and hydrodynamical equations can be solved analytically, and one 
finds that the small perturbations in energy density, velocity, etc are 
generated uniquely by the curvature profile, which is then the only 
quantity that needs to be specified. In \cite{Musco2} (hereafter referred 
to as Paper 2) the numerical scheme developed in Paper 1 was modified so 
as to use initial data coming from this quasi-homogeneous solution.

The present paper is continuing this line of work, probing much more 
deeply into the nearly-critical regime, using an updated version of our 
previous computer code implementing an adaptive mesh refinement scheme. We 
investigate the behaviour occurring in the nearly-critical regime with 
particular reference to a possible breaking of the scaling law.

After this introduction, Section 2 gives a brief summary of the equations 
used and of the method for specifying the initial conditions; Section 3 
describes the numerical method used for the simulations, with particular 
focus on the new modifications to include the adaptive mesh refinement. 
Results are then presented in Section 4 and Section 5 contains 
conclusions. Throughout, we use units for which $c = G = 1$ except where 
otherwise stated. In Appendix A, we list some typographical errors in 
Paper 2 which have come to light and in Appendix B, we present a 
convergence test of the code.

%=======================================================

%=======================================================
\section{Mathematical formulation of the problem}
\label{equations}
 For the calculations described here, we have followed the same basic 
methodology as described in Papers 1 and 2. We give a brief summary of it 
here; full details are contained in the previous papers.

We use two different formulations of the general relativistic hydrodynamic 
equations: one for setting the initial conditions and the other for 
studying the black hole formation. Throughout, we are assuming spherical 
symmetry and that the medium can be treated as a perfect fluid; we use a 
Lagrangian formulation of the equations with a radial coordinate $r$ which 
is co-moving with the matter.

For setting the initial conditions, it is convenient to use a diagonal 
form of the metric, with the time coordinate $t$ reducing to the standard 
FRW time in the case of a homogeneous medium with no perturbations. (This 
sort of time coordinate is therefore often referred to as ``cosmic 
time''). We write this metric in the form given by Misner \& Sharp (MS) 
\cite{Misner}, whose approach we follow here in writing the GR 
hydrodynamic equations:
 \beq 
ds^2=-a^2\,dt^2+b^2\,dr^2+R^2\lp d\theta^2+\sin^2\theta d\varphi^2\rp, 
\label{sph_metric}
 \eeq
 with the coefficients $a$, $b$ and $R$ being functions of $r$ and $t$ and 
$R$ playing the role of an Eulerian radial coordinate. Using the 
definitions
 \beq
D_t\equiv\frac{1}{a}\lp \frac{\partial}{\partial t}\rp, \label{D_t}
\eeq
\beq D_r\equiv\frac{1}{b}\lp
\frac{\partial}{\partial r}\rp, \label{D_r}
\eeq
 one defines the quantities
\beq
U \equiv D_t R, \label{U}
\eeq
and
\beq 
\Gamma \equiv D_r R,
\label{Gamma1}
\eeq
 where $U$ is the radial component of four-velocity in the ``Eulerian'' 
frame and $\Gamma$ is a generalized Lorentz factor. The metric coefficient 
$b$ can then be written as
 \beq
b \equiv \frac{1}{\Gamma}\frac{\partial R}{\partial r},     
\eeq
 For a radiative medium where rest-mass makes a negligible contribution to 
the energy density, the basic GR hydrodynamic equations can then be 
written in the following form (where we use the notation that $e$ is the 
energy density, $p$ is the pressure, $\rho$ is the compression factor, and 
$M$ is the mass contained inside radius $R$):
 \beq
D_tU=-\lP \frac{\Gamma}{(e+p)}D_rp+\frac{M}{R^2}+4\pi Rp\rP,
\label{Euler1} 
\eeq 
\beq D_t\rho=-\frac{\rho}{\Gamma
R^2}D_r(R^2U), \label{D_trho} 
\eeq 
\beq D_t
e=\frac{e+p}{\rho}D_t\rho,\label{D_te} 
\eeq 
\beq D_t M=-4\pi R^2 pU,
\label{D_tM} 
\eeq 
\beq D_r a=-\frac{a}{e+p}D_r p,\label{D_ra} 
\eeq
\beq D_r M=4\pi R^2 \Gamma e \, , \label{D_rM}
\eeq 
plus a constraint equation
\beq
\Gamma^2=1+U^2-\frac{2M}{R}\, .
\label{Gamma} 
\eeq
 We also have the equation of state for a radiative fluid:
 \beq
p = \textstyle{\frac{1}{3}}e\,.
\label{eq_state}
\eeq

The initial conditions are set by introducing a perturbation of the 
otherwise uniform medium representing the cosmological background 
solution, with the length-scale of the perturbation $R_0$ being much 
larger than the cosmological horizon $R_H\equiv H^{-1}$. Under these 
circumstances, the perturbations in $e$ and $U$ can be extremely small 
while still giving a large amplitude perturbation of the metric (as is 
necessary if a black hole is eventually to be formed) and the above system 
of equations can then be solved analytically to first order in the small 
parameter $\epsilon \equiv (R_H/R_0)^2 << 1$. A full discussion of this 
has been given in Paper 2. The result, called the ``quasi-homogeneous 
solution'', gives formulae for the perturbations of all of the metric and 
hydrodynamical quantities in terms only of a curvature perturbation. When 
$\epsilon<<1$, the curvature perturbation commonly denoted by $\zeta$ is 
time-independent \cite{Lyth}, and it is convenient to represent it in 
terms of another quantity $K(r)$ similar to the constant $K$ appearing in 
the standard form of the FRW metric but now depending on $r$. The 
perturbed form of the metric coefficient $b$ is then written as
 \beq
b = \frac{R^\pr}{\sqrt{1-K(r)r^2}}(1+\epsilon\tilde b)\, ,
 \label{b_def}
\eeq
 where here, and in the following, the tilde denotes a perturbation 
quantity and the prime denotes a derivative with respect to $r$. (Note 
that throughout this discussion of the initial conditions, we are working 
to first order in $\epsilon$, but in our computations of the subsequent 
evolution we use the full set of hydrodynamical equations with no 
linearization.) The energy density and all of the other variables of the 
MS equations are perturbed in the following way, which we show explicitly 
for the energy density:
 \beq
e =  e_b(1+\epsilon \tilde e)
\label{perts}
\eeq
 where $e_b$ represents the background unperturbed solution and $\tilde e$ 
gives the perturbation profile. Inserting the perturbed quantities into 
(7) - (13) and performing a linearization in $\epsilon$, one can obtain 
time-independent analytical solutions for all of the tilde-quantities 
\footnote{As explained in Paper 2, the quasi-homogeneous solution holds 
also for multiple fluids; in that case, the tilde-quantities depend on 
both $r$ and $t$.}. With the equation of state (\ref{eq_state}), $\tilde 
e$ is given by
 \beq
\tilde e = \frac{2}{3} \frac{1}{3r^2}\lP r^3K(r)\rP^\pr r_0^2
\label{e_tilda}    
\eeq
 where $r_0$ is the comoving length-scale of the overdense region of the 
perturbation, determined by the particular expression used for $K(r)$. The 
time variation within the quasi-homogeneous solution is given by 
$\epsilon(t)$, which follows the standard relation for a cosmological 
growing mode (in the present case, $\epsilon \propto t$).

To characterize the amplitude of the perturbation, we use the integrated 
quantity $\delta$, which measures the relative mass excess within the 
overdense region, as frequently done in the literature. In the 
quasi-homogeneous solution, this is given by
 \beq 
\delta(t) \equiv \lp \frac{4}{3}\pi r_0^3\rp^{\!-1} 
\displaystyle{\int_0^{r_0} 4\pi \frac{e-e_b}{e_b}r^2dr} \ug 
\frac{2}{3}\epsilon(t)K(r_0)r_0^2
\label{delta_s} 
\eeq
 which also has the familiar linear growth with cosmic time. For 
discussing the scaling law, we need to have a measure of the perturbation 
amplitude defined consistently for different cases, and for this we use 
$\delta$ as given by expression (\ref{delta_s}) with $\epsilon$ set equal 
to $1$. This gives very similar values to those actually calculated at 
horizon crossing, with non-linear evolution taken into account, but it can 
be calculated with greater precision since it is given analytically.
 
In the present work, we make the choice of using a Gaussian profile for 
$K(r)$, normalized to be equal to $1$ at the centre:
 \beq
K(r) = \exp{\lp-\frac{r^2}{2\Delta^2}\rp}\,.
\label{curv_profile}
\eeq
 Substituting this into (\ref{e_tilda}), one obtains 
 \beq
\tilde{e}  \ug  \frac{2}{3} r_0^2 \lP 1 \,-\, \frac{r^2}{3\Delta^2} \rP
\exp{\lp -\frac{r^2}{2\Delta^2}\rp}\,.
\label{e_tilda2}
\eeq
 Since $r = r_0$ at the outer edge of the overdensity (where $\tilde{e} 
\to 0$), one then has $r_0^2 = 3\Delta^2$ and so (\ref{e_tilda2}) can be 
rewritten as
 \beq
\tilde{e}  \ug  2 \Delta^2 \lP 1 \,-\, \lp\frac{r}{r_0}\rp^{\!2} \rP 
\exp{\lp -\frac{3}{2}\lp \frac{r}{r_0}\rp^{\!2}\rp}\,,
\label{e_tilda3}
\eeq
 or
 \beq
\tilde{e}  \ug  2 \Delta^2 \lP 1 \,-\, \lp\frac{R_b}{R_0}\rp^{\!2} \rP
\exp{\lp -\frac{3}{2}\lp \frac{R_b}{R_0}\rp^{\!2}\rp}\,,  
\label{mex_hat}
\eeq
 in terms of the radial coordinate $R_b$ of the unperturbed background 
solution, given by $R_b=s(t)r$ where $s(t)$ is the cosmological scale 
factor. This expression is that for the well-known mexican-hat profile, as 
used in previous calculations by Niemeyer \& Jedamzik \cite{Jedamzik1} and 
ourselves \cite{Musco1, Musco2}. Inserting the expression for $r_0$ into 
(\ref{delta_s}) gives
 \beq
\delta = 2\Delta^2 \exp\left(-\textstyle{\frac{3}{2}}\right)
\eeq
 when $\epsilon=1$ and so the perturbation amplitude can be characterized 
alternatively by $\Delta$. In Papers 1 and 2, the $\delta_c$ for this 
profile was found to be $\simeq 0.45$, which corresponds to $\Delta_c 
\simeq 1$. For further details about setting initial conditions using the 
quasi-homogeneous solution, the reader is referred to Paper 2.

\vspace{0.5cm}

The MS approach using cosmic time slicing is convenient for setting 
initial conditions but has a well-known drawback for calculations of black 
hole formation in that singularities are formed rather quickly when using 
it and then further, potentially observable, evolution cannot be followed 
unless an excision procedure is used. Various slicing conditions can be 
used to avoid this difficulty but for calculations in spherical symmetry 
it is particularly convenient to use null slicing. In our work we have 
used the ``observer time'' null-slicing formulation of Hernandez \& Misner 
\cite{Hernandez} where the time coordinate is taken as the time at which 
an outgoing radial light ray emanating from an event reaches a distant 
observer. (In the original formulation, this observer was placed at future 
null infinity but for calculations in an expanding cosmological background 
we use an FRW fundamental observer sufficiently far from the perturbed 
region so as to be unaffected by the perturbation.) We use the MS approach 
for setting up initial data and for evolving it so as to produce data on a 
null slice. This is then used as input for our observer-time code with 
which we follow the black hole formation.

For the observer-time calculation, the metric (\ref{sph_metric}) is 
re-written as 
 \beq 
ds^2=-f^2\,du^2-2fb\,dr\,du+R^2\left(d\theta^2+\sin^2\theta 
d\varphi^2\right), \label{nullmetric} \eeq
 where $u$ is the observer time and $f$ is the new lapse function. The 
operators equivalent to (\ref{D_t}) and (\ref{D_r}) are now
 \beq
D_t \equiv \frac{1}{f}\left(\frac{\partial}{\partial u}\right),
\eeq
 \beq
D_k \equiv \frac{1}{b}\left(\frac{\partial}{\partial r}\right) ,
\eeq
 where $D_k$ is the radial derivative in the null slice and the 
corresponding derivative in the Misner-Sharp space-like slice is given by
 \beq
D_r=D_k-D_t.
\eeq
 The hydrodynamic equations can then be formulated in an analogous way to 
what was done in cosmic time. We will not repeat this here but refer the 
reader to our discussion in \cite{Musco1}.

%=======================================================

%=======================================================
\section{The method of calculation}
\label{method}

As in our previous work, the present calculations have been made with an 
explicit Lagrangian hydrodynamics code based on that of Miller \& Motta 
(1989) \cite{Miller1} but with the grid organized in a way similar to that 
in the code of Miller \& Rezzolla (1995) \cite{Miller2} which was designed 
for calculations in an expanding cosmological background. The code has a 
long history and has been carefully tested in its various forms. Full 
details of the methods used have been given in the above papers, in Papers 
1 and 2, and in other papers cited there. Some further information about 
tests of the code, including a new convergence test, is given in Appendix 
B.

The code uses a staggered grid and the co-moving coordinate is a mass-type 
coordinate which we will refer to as $\mu$ in the discussion below. The 
basic grid has logarithmic spacing, allowing it to reach out to very large 
radii while giving finer resolution at small radii. The initial data for 
our present calculations was derived from the quasi-homogeneous solution 
and was specified on a space-like slice (at constant cosmic time) with 
$\epsilon = 10^{-2}$, giving $R_0 = 10\,R_H$. The outer edge of the grid 
was placed at $90\,R_H$, which is far enough away so that there is no 
causal contact between it and the perturbed region during the time of the 
calculations. The initial data was then evolved using the MS equations 
(\ref{Euler1}-\ref{Gamma}), so as to generate a second set of initial data 
on a null slice (at constant observer time). To do this, an outgoing 
spherical light pulse was traced out from the centre, and parameter values 
were noted as it passed each grid zone. It is not necessary to continue to 
evolve the solution significantly behind the wave front in this part of 
the calculation, since that region (within which singularities might form) 
does not have causal contact with the wave front. The null-slice initial 
data, constructed in this way, was then evolved using the observer-time 
equations \cite{Musco1}. From here on, we will refer to the observer-time 
coordinate as $t$ rather than $u$.

In the present work, our aim was to study PBH formation for much smaller 
values of $(\delta-\delta_c)$ than was possible with our previous version 
of the code. The particular problem which needed to be overcome is that as 
the PBH is forming, a semi-void appears around it and this becomes 
progressively deeper as $\delta$ gets closer to $\delta_c$. For a 
Lagrangian code with a co-moving grid, this means that the Eulerian widths 
of zones in the semi-void region become very large, producing truncation 
errors that crash the code. The situation is made worse by the fact that a 
highly-relativistic wind blows matter across this region and this needs 
good resolution in order to be well-represented. As mentioned above, the 
basic code was already using a logarithmically-spaced grid in order to be 
able to simultaneously treat both the small scales on which the PBH was 
forming and also the large scales necessary for representing the 
surrounding expanding universe up to scales much larger than the 
cosmological horizon. For dealing with the problem of the semi-void 
region, we introduced an adaptive mesh refinement scheme (AMR) on top of 
the existing logarithmic grid. This has allowed us to follow PBH formation 
for values of $(\delta - \delta_c)$ up to eight orders of magnitude 
smaller than we could do before.

Our AMR uses a fully-threaded-tree algorithm, making its refinement on the 
basis of the Eulerian width of the zone $\Delta R$ (which then, in turn, 
impacts on the truncation error). The outline of the procedure is as 
follows. Every 20 time-steps, a check is made to see whether the 
fractional zone spacing $\Delta R/R$ is greater than 10\% for any zone. If 
it is, that zone is subdivided introducing a new grid point at its centre 
(in terms of the co-moving radial coordinate $\mu$) and quantities at the 
new grid-point and in the two newly-created zones to either side of it are 
calculated by means of cubic interpolation. Only one zone is subdivided at 
a time, allowing the solution to relax before performing any subsequent 
sub-division. Doing this, we have not found it necessary to introduce any 
artificial viscosity into the code for suppressing numerical noise, an 
important point for interpreting the appearance or absence of shocks. Also 
every 20 time steps (but offset by 10 time steps), we carry out a check to 
see whether $\Delta R/R$ has become less than 4\% across any zone 
resulting from a previous subdivision. If so, then it is merged with the 
zone from which it was previously separated if that zone has the same 
$\Delta \mu$. Again, only one merging is carried out at a time to allow 
relaxation. Other grid-management routines are used so as to maintain a 
good overall grid structure.

We have successfully used the scheme with more than thirty levels of 
refinement and its performance can be judged from the results which we 
present in the next section. All of the features shown have been fully 
resolved with the AMR.

%===========================================================

%===========================================================

\section{Description of the results}

%%%%%%%%%%%%%%%%%%%%%%%%PICTURE%%%%%%%%%%%%%%%%%%%%%
\begin{figure}[ht!]
\centering
\includegraphics[width=8cm]{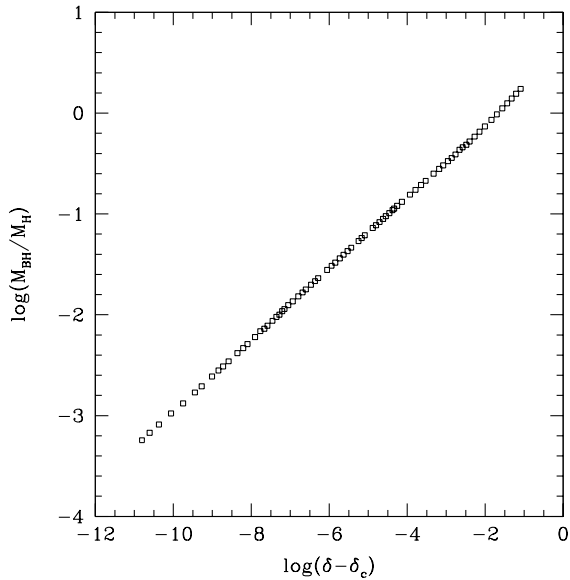}
 \caption{\label{fig.1} \small Scaling behaviour for $M_{BH}$ as 
function of $(\delta-\delta_c)^\gamma $ calculated for a radiative 
perfect fluid. For $M_{BH}\lesssim M_H$, the points are well-fitted by 
a scaling law with $\gamma = 0.357$ and $K=4.02$.}
 \end{figure}
%%%%%%%%%%%%%%%%%%%%%%%%%%%%%%%%%%%%%%%%%%%%%%%%%%
 In this section we discuss the numerical results which we have obtained 
for PBH formation in the radiative era, going closer to $\delta = 
\delta_c$ than we have done before. The main result is shown in figure 
\ref{fig.1} where the log of the black-hole mass $M_{BH}$ normalized with 
$M_H$ (the mass inside the cosmological horizon at the moment of horizon 
crossing) is plotted against the log of $(\delta-\delta_c)$. Note that the 
cosmological horizon mass continues to increase after the horizon-crossing 
time, and so it is not inconsistent to have $M_{BH} > M_H$.

We find that for $M_{BH}\lesssim M_H$ (which corresponds to 
$(\delta-\delta_c)\lesssim 2\cdot10^{-2}$), the points for different cases 
rather accurately follow a scaling law
 \beq
\frac{M_{BH}}{M_H} = K(\delta-\delta_c)^\gamma\,
\label{scaling}
\eeq
 with $\gamma = 0.357$ -- close to the expected value for a perfect fluid 
of radiative particles \cite{Evans}. For larger values of the masses, the 
curve steepens slightly, as seen also in the past \cite{Jedamzik1, Hawke, 
Musco1} but there is no sign of divergence from the scaling law at lower 
masses, even though we have now covered a range of masses in the 
scaling-law region of more than one thousand. The most extreme case shown 
has $(\delta-\delta_c)\sim10^{-11}$ and gives a black hole mass of $\sim 5 
\cdot 10^{-4}M_H$. These results are in contrast with those seen by Hawke 
\& Stewart \cite{Hawke}, where the scaling-law behaviour persisted for a 
range of masses of $\sim 100$ but terminated at low masses with the curve 
flattening out at a minimum value of $M_{BH}/M_H$. However, we want to 
stress that great care needs to be taken in making a comparison.

Firstly, in order to compare our results in figure \ref{fig.1} 
quantitatively with those in figures 7 \& 8 of \cite{Hawke}, we need to 
relate the two different mass normalizations used. The value of $M_H$ 
appearing in (\ref{scaling}) depends on the moment when it is measured. In 
common with much of the literature, we have evaluated it at the time of 
horizon crossing, whereas in \cite{Hawke} the perturbation was started 
already well inside the horizon and $M_H$ was evaluated at that initial 
time. Making a direct comparison is then difficult. Also, the profiles 
used for the perturbations were quite different. To get some idea, one can 
compare the upper part of the plots, where one sees a deviation away from 
the scaling law at large masses. In \cite{Hawke}, this deviation occurs at 
$M_{BH} \sim 0.1 \, M_H$ while we observe it occurring at $M_{BH}\sim 
M_H$. This indicates that the results in \cite{Hawke} probably need to be 
rescaled upwards in mass by about one order of magnitude in order to make 
the comparison (this has been confirmed in discussion with one of the 
previous authors), putting their plateau at $\sim 10^{-2.5}$ in our units, 
with a divergence from the scaling behaviour at the low-mass end beginning 
at $(\delta-\delta_c)\sim10^{-8}$ using our measure of perturbation 
amplitude. Their explanation for the minimum mass seen in their 
calculations was related to the occurrence of strong shocks in the 
lower-mass cases. With our initial conditions, we do not see these shocks 
(hence the continuation of the scaling law) but if we impose more general 
non-linear initial conditions within the cosmological horizon scale, then 
we {\em do} often see shocks which are consistent with those reported in 
\cite{Hawke} (although our code is not equipped to handle the strong shock 
conditions which they saw and so we can see only the beginnings of this 
behaviour). We therefore attribute the difference in results to the 
different initial conditions used.

In our simulations, we start with supra-horizon-scale perturbations 
derived from the quasi-homogeneous solution, which are still well within 
the linear regime as far as the hydrodynamical variables are concerned 
(although they are non-linear in the curvature, as mentioned previously). 
These perturbations, with only a growing component, have a very special 
relationship between the density profile and the velocity field. If we 
imposed as our initial conditions, at the same supra-horizon-scale, a 
perturbation of density but not of velocity (as we did in \cite{Musco1}), 
then the perturbation would reconfigure as it evolved, going back to this 
special relationship between the density profile and the velocity field 
before horizon crossing. (Comparing with a similar perturbation which is 
linear in {\em all} quantities, including the curvature, one can think 
there of it as being a superposition of growing and decaying modes: as it 
evolves, the decaying mode dies away, leaving only the growing mode.) When 
this special type of perturbation becomes non-linear in the fluid 
quantities as well as in the curvature, it still holds together and does 
not produce shocks in collapses producing black holes, as a general 
perturbation set up at late times can do. Shibata \& Sasaki \cite{Shibata} 
who used a similar type of initial perturbation to ours, but formulated in 
a completely different way, also did not mention seeing shock formation in 
their collapses producing black holes.

%%%%%%%%%%%%%%%%%%%%%%%%PICTURE%%%%%%%%%%%%%%%%%%%%%
\begin{figure}[ht!]
\centering
\includegraphics[width=6cm]{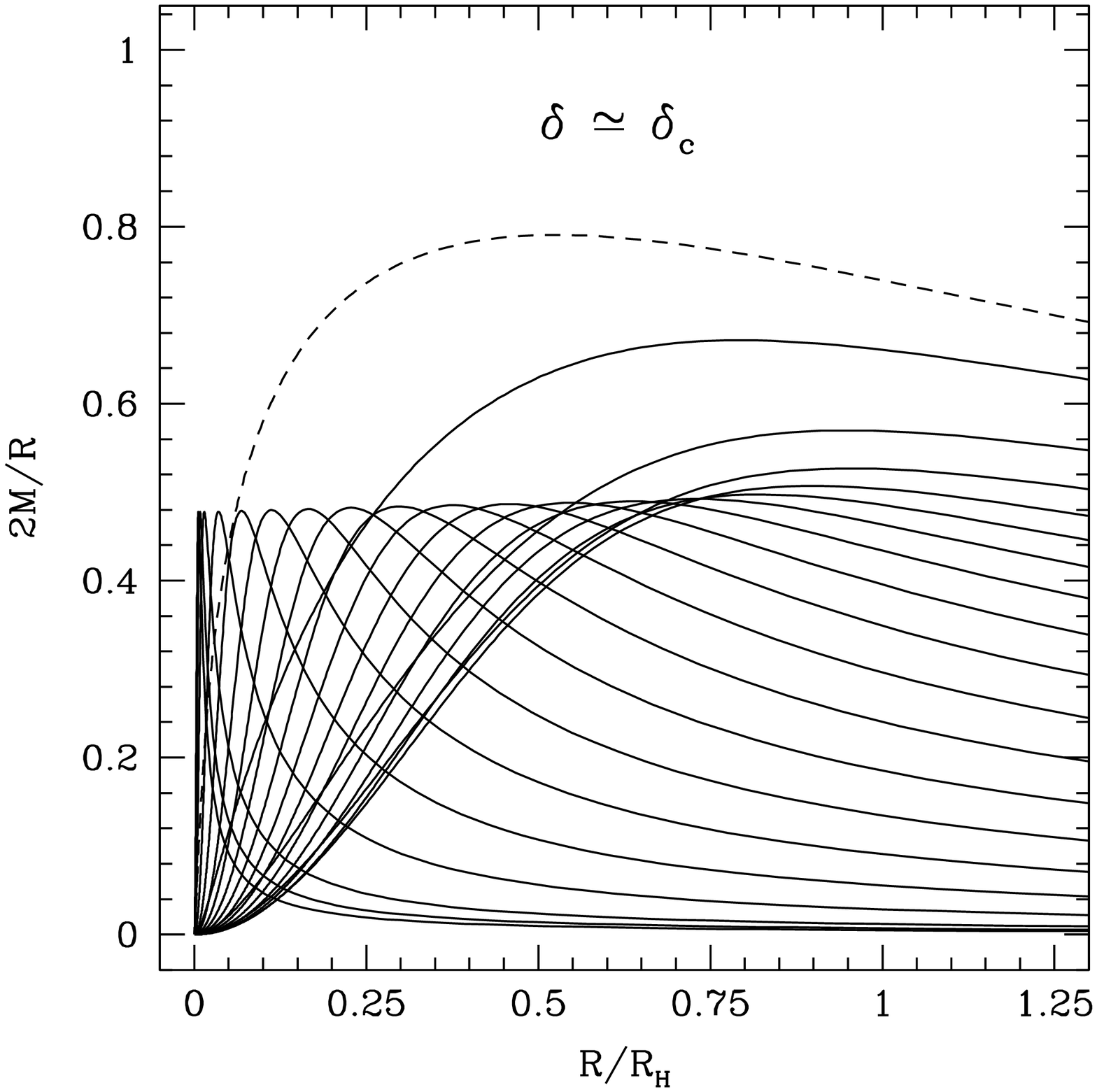}
\includegraphics[width=6cm]{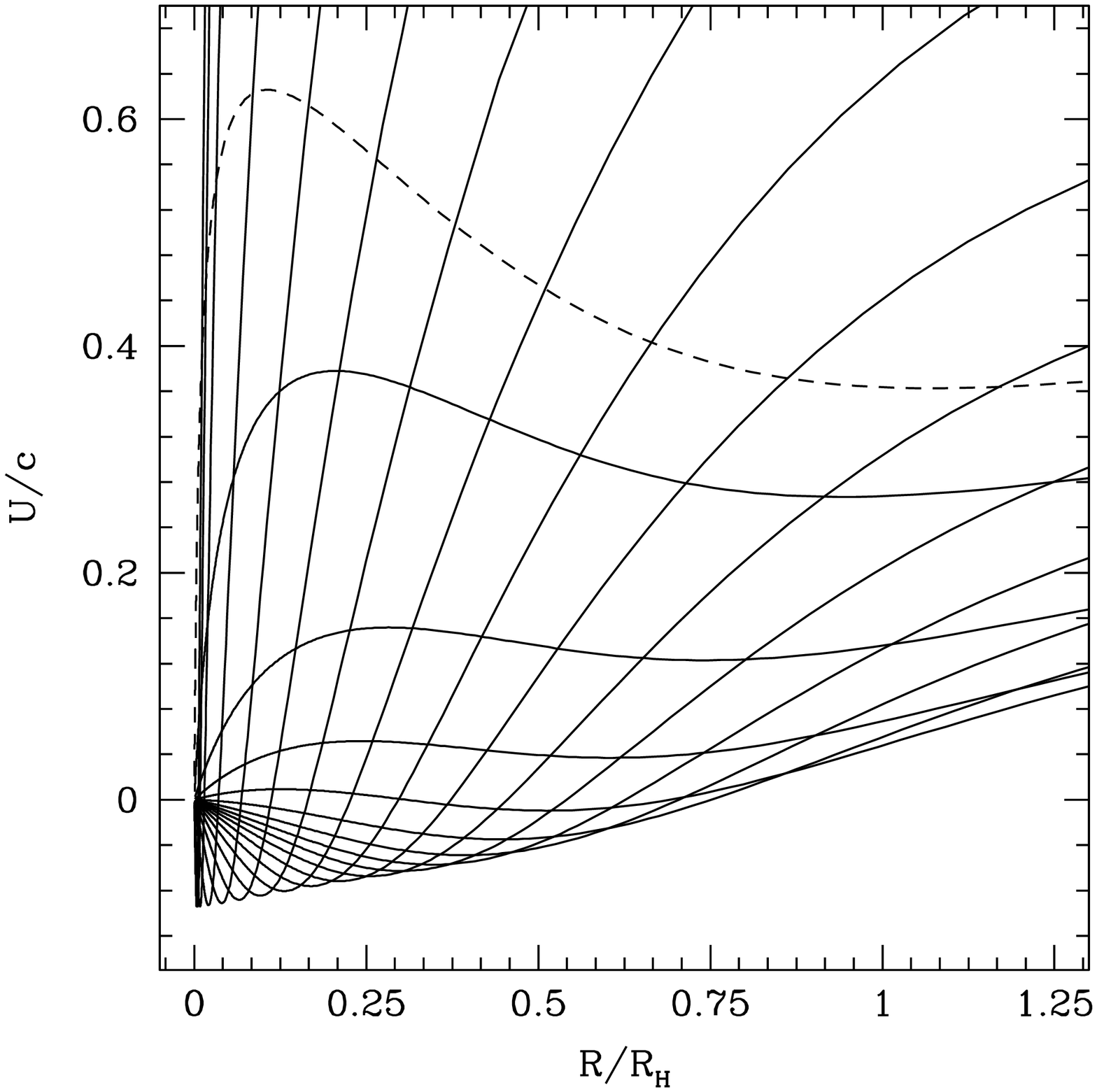}
\includegraphics[width=6cm]{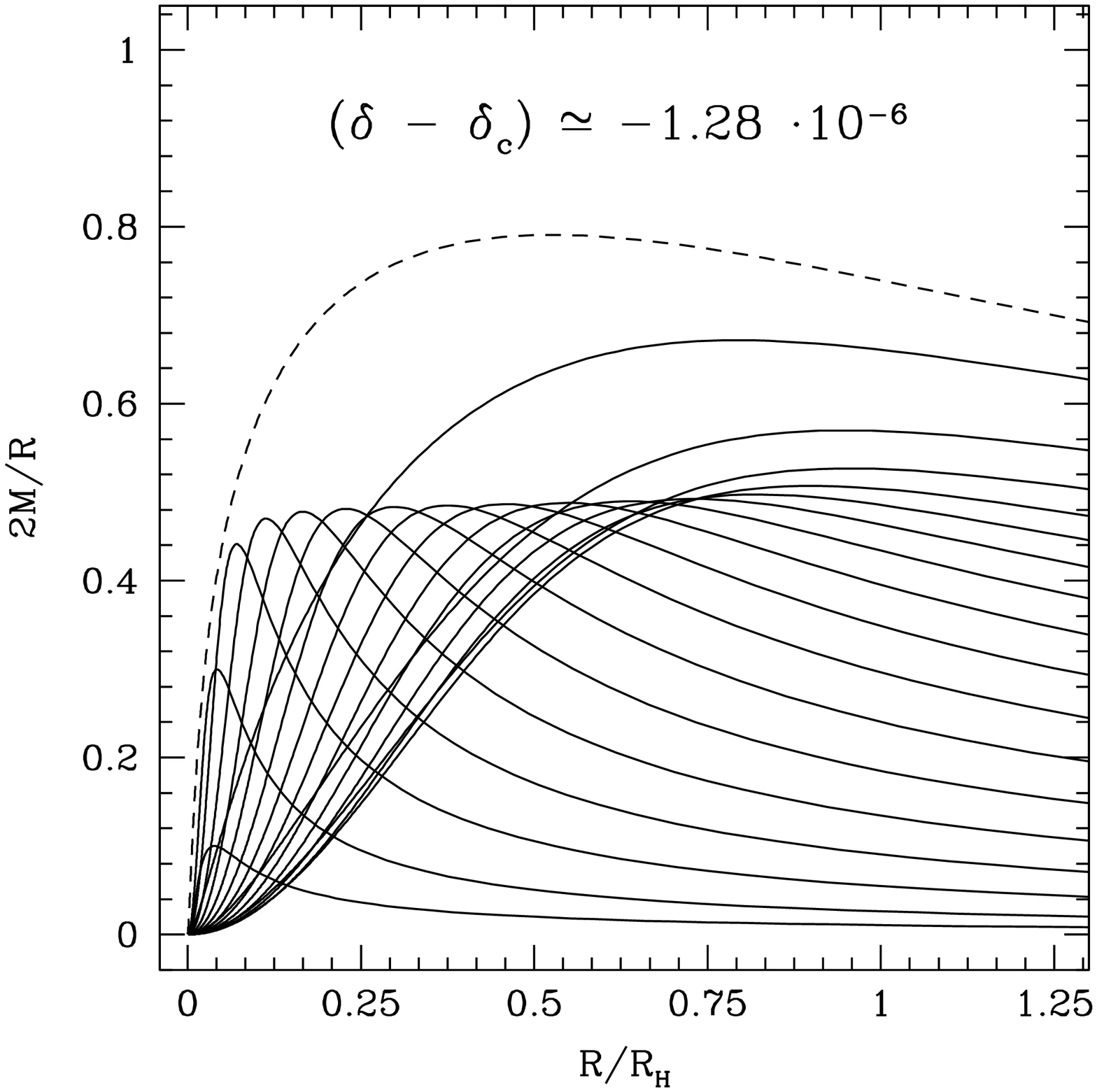}
 \caption{\label{fig.2}\small The top two frames show the behaviour of 
$2M/R$ and the radial four-velocity $U$ for a nearly-critical case 
($\delta \simeq \delta_c)$, plotted against $R/R_H$ at different time 
levels, with the dashed curve representing the initial conditions used 
by the observer-time code. $R_H$ is the horizon scale at the moment of 
horizon crossing. The bottom frame shows the equivalent plot of $2M/R$ 
for a sub-critical case with $(\delta - \delta_c) \simeq - 1.28 \cdot 
10^{-6}$ which does not produce a black hole. Some information about 
timing is given later, in figure \ref{fig.5}.}
 \end{figure}
%%%%%%%%%%%%%%%%%%%%%%%%%%%%%%%%%%%%%%%%%%%%%%%%%%

We next present some further results from our simulations of PBH formation 
near to the critical limit which give further insight into the nature of 
the process. The critical case $(\delta = \delta_c)$ separates cases which 
collapse to give a black hole from ones which bounce and merge back into 
the ambient medium. In figure \ref{fig.2} we show some plots for a 
nearly-critical case and a subcritical one. The top two frames show 
results from a run which was too close to critical for us to tell whether 
it would produce a black hole or not by the time that it terminated. Here 
one can see an important characteristic feature of nearly-critical 
collapse. The quantity $2M/R$ is a key diagnostic of the behaviour, going 
to 1 at horizons, both the cosmological horizon and the apparent horizon 
of the forming black hole. In the top left-hand frame, $2M/R$ is plotted 
as a function of $R/R_H$ at different time levels, where $R_H$ is the 
horizon scale at the moment of horizon crossing. The dashed curve 
represents the initial conditions used by the observer-time code. This 
plot is showing just the inner part of the solution where the collapse 
occurs: the outer part where $2M/R$ rises to 1 again at the cosmological 
horizon is at far larger radii. As time proceeds, the maximum of $2M/R$ 
initially decreases to around $0.5$, moving outwards in radius, but then 
it moves towards the centre, maintaining an almost constant value but with 
a very slow decrease towards an eventual value of $\sim 0.48$. This marks 
a ``critical surface'' that separates cases giving collapse to a black 
hole from ones which do not \cite{Gundlach}. Super-critical and 
sub-critical cases eventually deviate away from this. The right-hand frame 
shows the corresponding behaviour of the radial four velocity $U$ (note 
that these profiles look rather different when viewed in observer-time 
from the familiar picture in cosmic-time). Initially the velocity is 
positive everywhere with the general cosmological expansion just being 
slowed somewhat in the region of the perturbation. As time goes on, the 
expansion is progressively slowed down both because the universe as a 
whole is decelerating and because the effect of the perturbation becomes 
progressively greater in the central regions. After the dashed curve 
marking the initial time, the maximum in the velocity profile becomes 
progressively less pronounced and eventually the expansion reverses into a 
collapse in the central regions. As time proceeds further, the maximum 
infall velocity increases but the size of the collapsing region shrinks, 
eventually tending towards zero. Outside the collapsing region, one sees 
increasing positive velocities, representing a wind which takes matter 
away from the central condensation. We will analyse this in more detail 
when we consider a super-critical collapse which produces a black hole. 
During the ``equilibrium phase'', when the maximum of $2M/R$ (which from 
now on we will refer to as $(2M/R)_{peak}$) remains roughly constant, the 
location of this maximum corresponds to that where the velocity goes to 
zero, separating the collapsing region from that of the outgoing wind. At 
the location of $(2M/R)_{peak}$, there is a continuing close balance 
between gravitational forces and pressure gradients which is a key feature 
of critical collapse. As the collapsing region shrinks, a balance is kept 
all the way down to zero size in the special critical case $\delta = 
\delta_c$. For sub-critical cases, with $\delta<\delta_c$, pressure 
eventually wins over gravity: the remaining material in the collapsing 
region bounces, with a consequent decrease of $(2M/R)_{peak}$ (see the 
bottom frame of figure \ref{fig.2}), and disperses into the surrounding 
medium. For nearly-critical cases, this is a surprisingly violent process, 
as has been described in \cite{Musco1}. For super-critical cases, with 
$\delta>\delta_c$, gravity eventually wins over pressure, $(2M/R)_{peak}$ 
then grows towards 1 and a black hole is formed. The rest of the present 
section contains a detailed discussion of some of the features of this, as 
seen in the simulations.

%%%%%%%%%%%%%%%%%%%%%%%%PICTURE%%%%%%%%%%%%%%%%%%%%%
\begin{figure}[ht!]
\centering
\includegraphics[width=6cm]{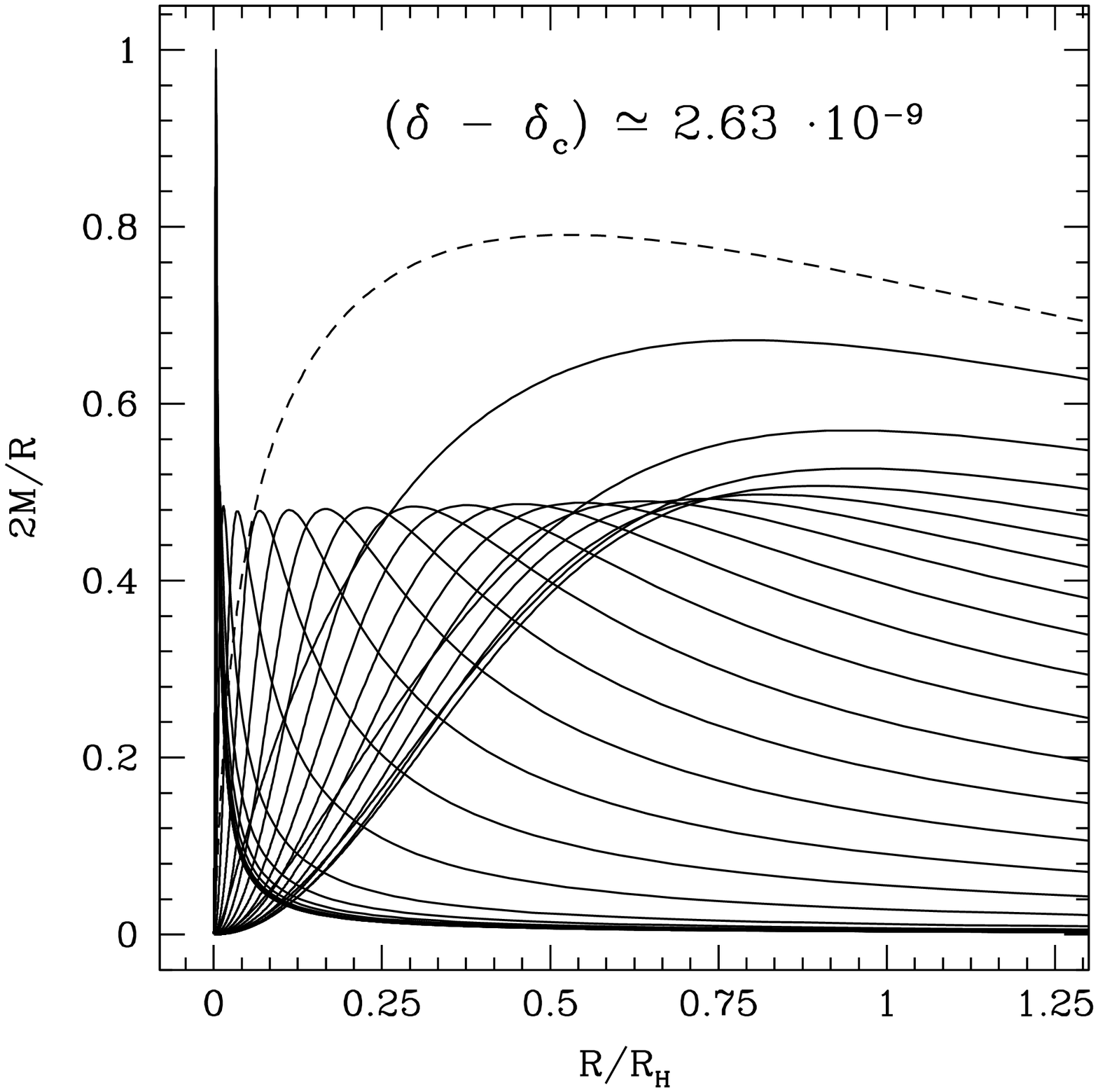}
\includegraphics[width=6cm]{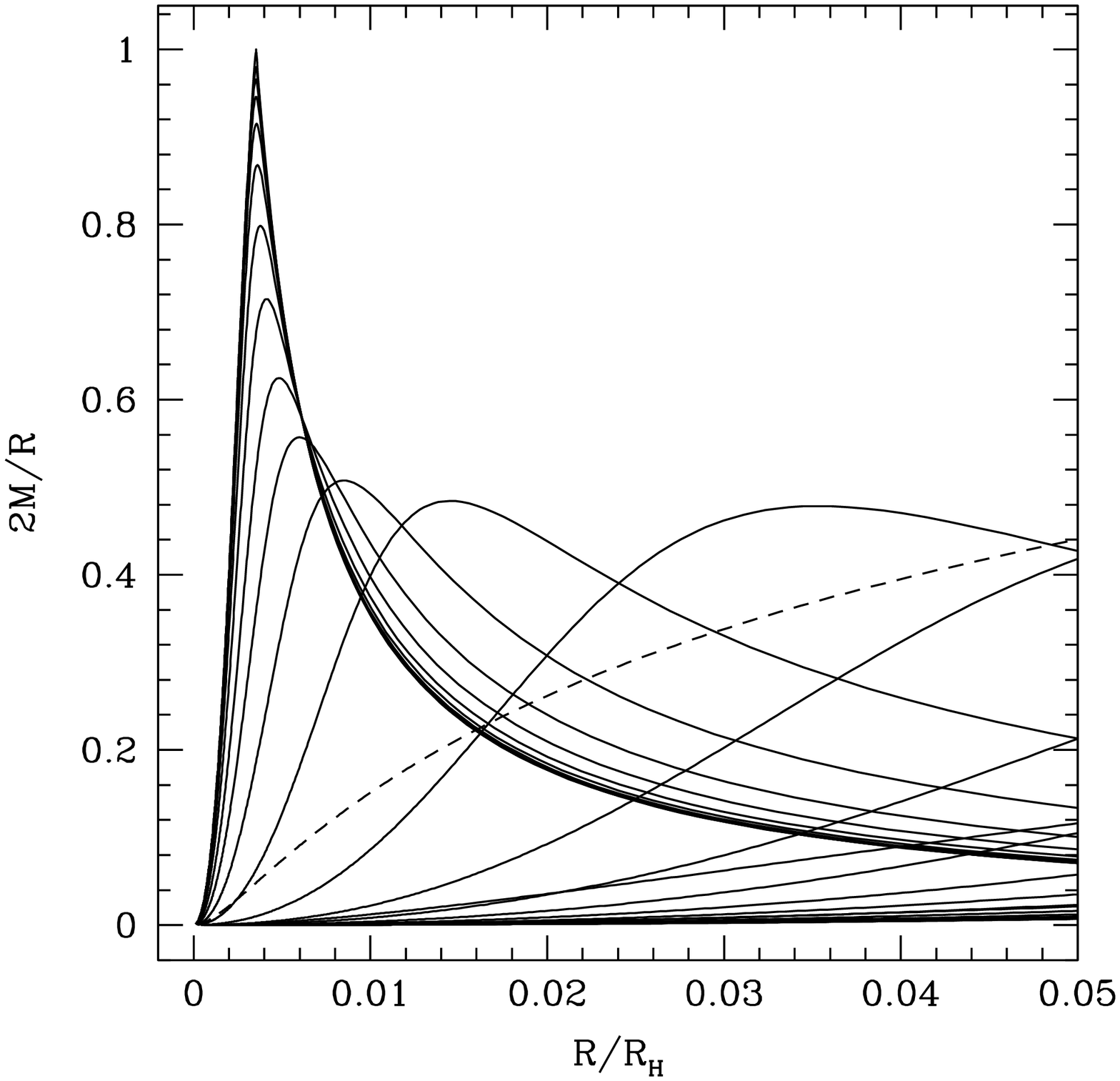}
\includegraphics[width=6cm]{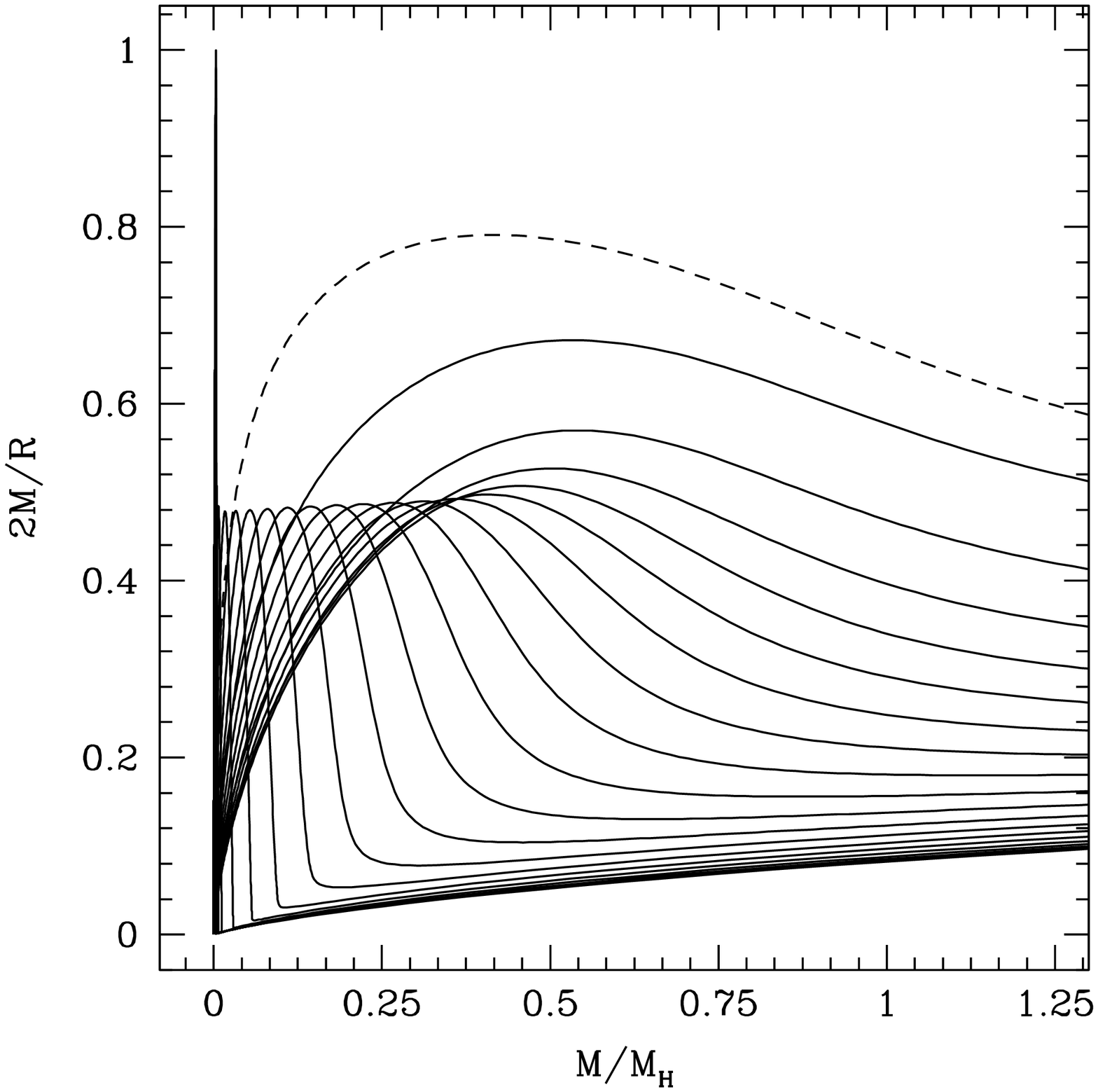}
\includegraphics[width=6cm]{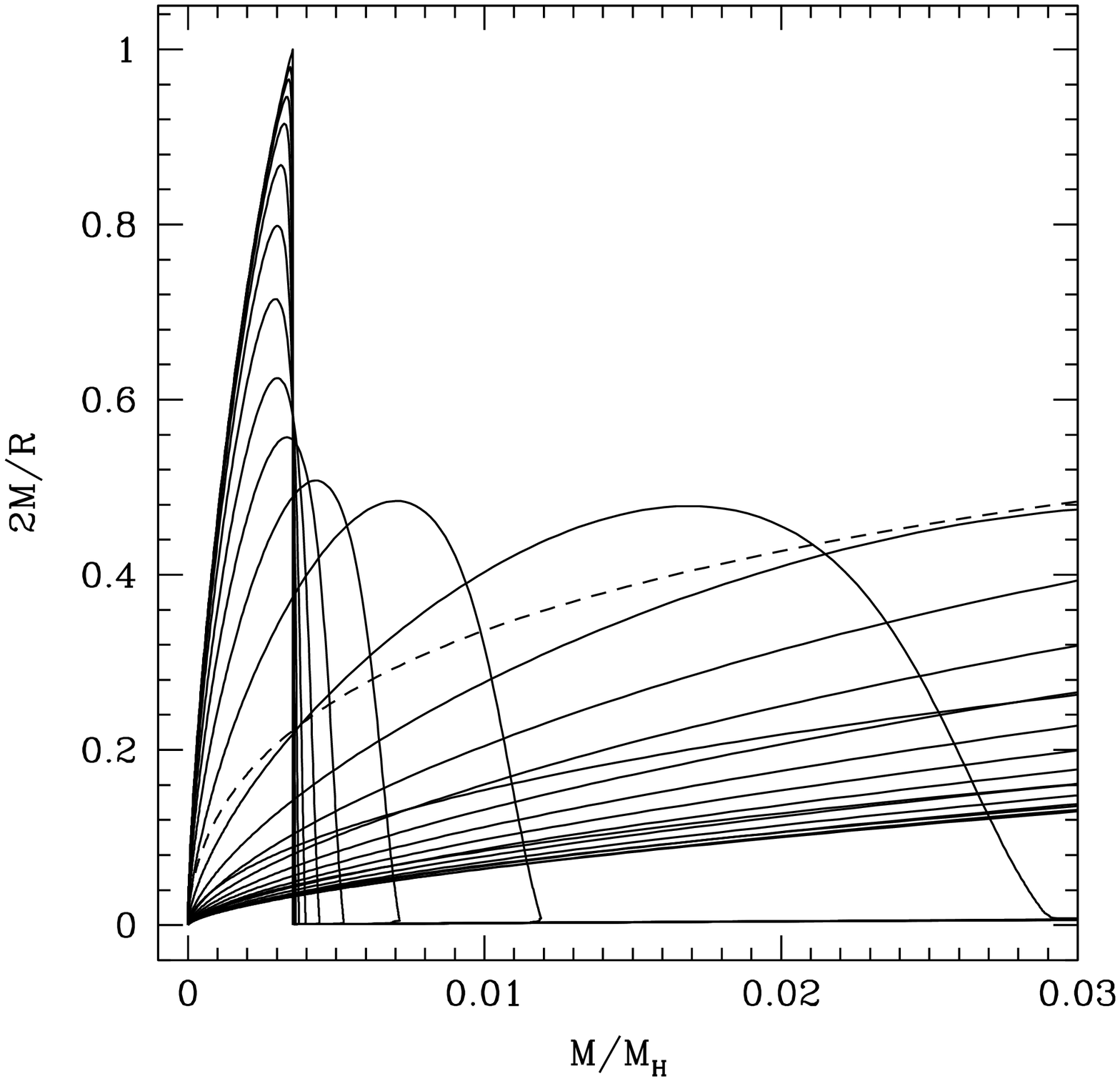}
 \caption{\label{fig.3}\small These plots, for a representative but 
fairly extreme case where a black hole is formed, show different views 
of the behaviour of $2M/R$ as a function of $R/R_H$ or $M/M_H$ at 
different time levels with the dashed curve representing the initial 
conditions used by the observer-time code. The perturbation has 
$(\delta - \delta_c) \simeq 2.63 \cdot 10^{-9}$ and the collapse gives 
rise to a black hole with a mass $M_{BH} \simeq 3.54 \cdot 10^{-3} 
M_H$. In the top two frames, $2M/R$ is plotted against $R/R_H$ with 
the right-hand frame being an enlargement of the inner parts of the 
left-hand one. The bottom two frames show the same data plotted as a 
function of $M/M_H$. Some information about timing is given later, in 
figure \ref{fig.5}.}
 \end{figure}
%%%%%%%%%%%%%%%%%%%%%%%%%%%%%%%%%%%%%%%%%%%%%%%%%%

For this discussion of super-critical collapse, we will focus on a 
particular representative case which is typical of those fairly close to 
the critical limit. This case has $(\delta - \delta_c) \simeq 2.63 \cdot 
10^{-9}$ and forms a black hole with mass $M_{BH} \simeq 3.54 \cdot 
10^{-3} M_H$. Figure \ref{fig.3} shows various views of the $2M/R$ 
profiles. In the top two frames, $2M/R$ is plotted against $R/R_H$ with 
the right-hand frame being an enlargement of the inner parts of the 
left-hand one. The top left-hand frame is the counterpart of the plots 
which we have shown for the critical and sub-critical cases and one can 
see the ``intermediate state'' with its almost-constant value of 
$(2M/R)_{peak}$ and the eventual precipitous rise of this towards 1 as the 
black hole forms. (Note that in observer time, the black hole only fully 
forms asymptotically, when the time as measured by a distant observer 
tends to infinity, but we terminated the calculation when the central 
value of the lapse $f$ had fallen below $10^{-10}$.) The top right-hand 
frame is an enlargement of the central part, showing how the solution 
diverges away from the intermediate state and enters the final black-hole 
collapse phase. The bottom two frames show the same data plotted as a 
function of $M/M_H$. This is useful for showing in a clear way that when 
the collapsing region reaches the intermediate state, it then evolves by 
losing material to the outside while maintaining almost the same 
compactness until the final precipitous collapse. Again, the right-hand 
frame is an enlargement of the central part of the left-hand one, showing 
more detail of the departure from the intermediate state.

%%%%%%%%%%%%%%%%%%%%%%%%PICTURE%%%%%%%%%%%%%%%%%%%%%
\begin{figure}[ht!]
\centering
\includegraphics[width=6cm]{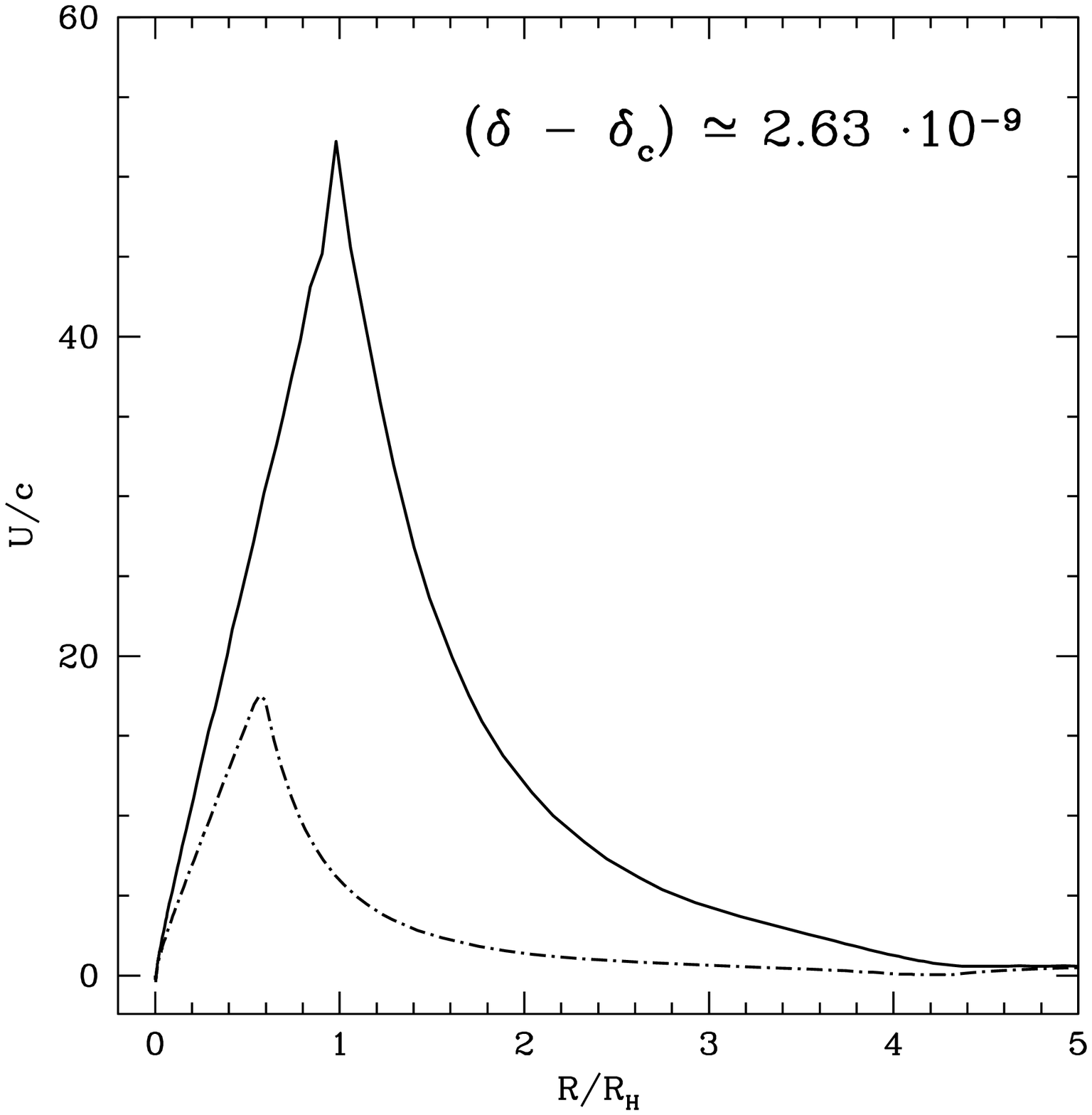}
\includegraphics[width=6cm]{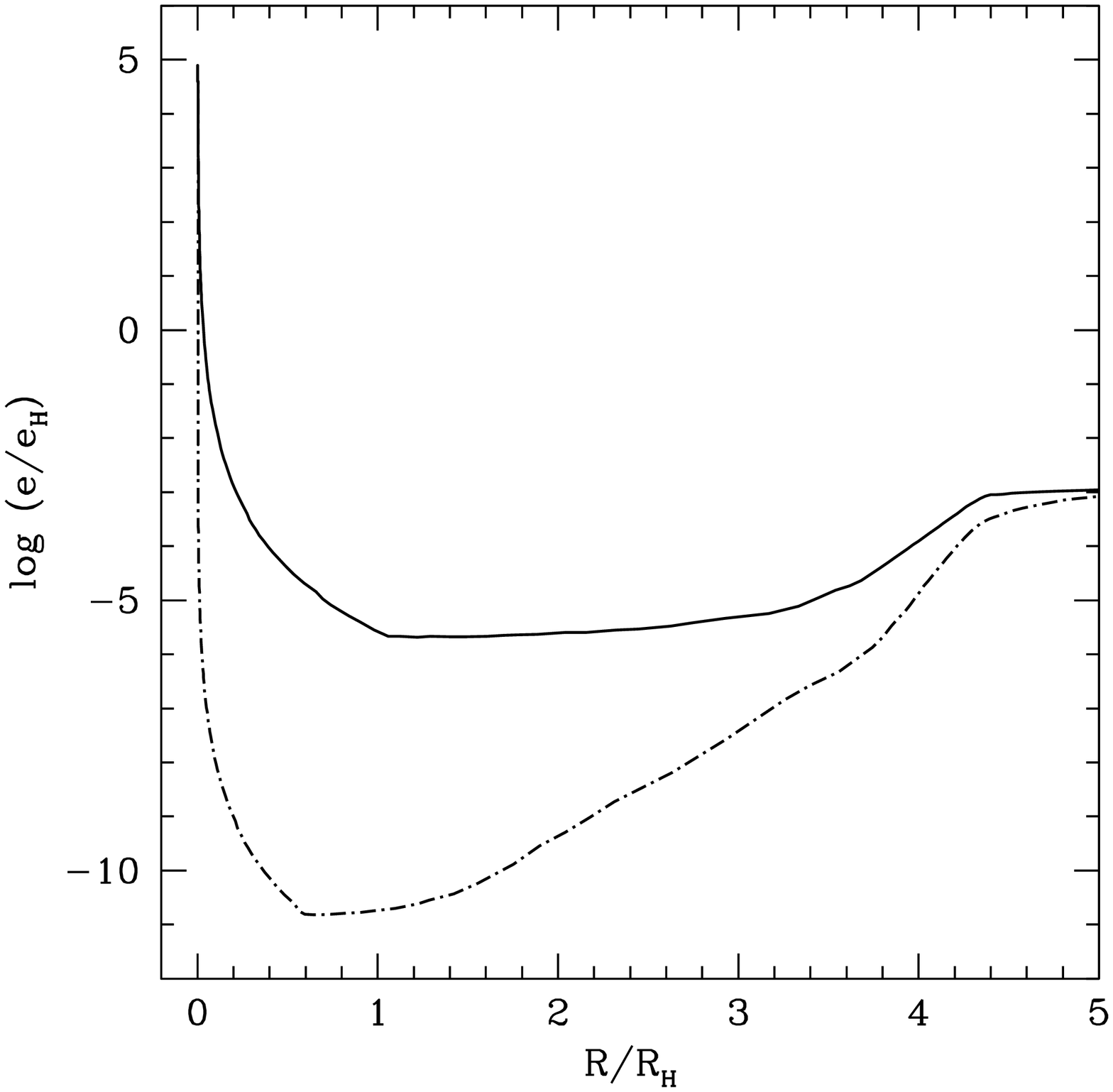}
\includegraphics[width=6cm]{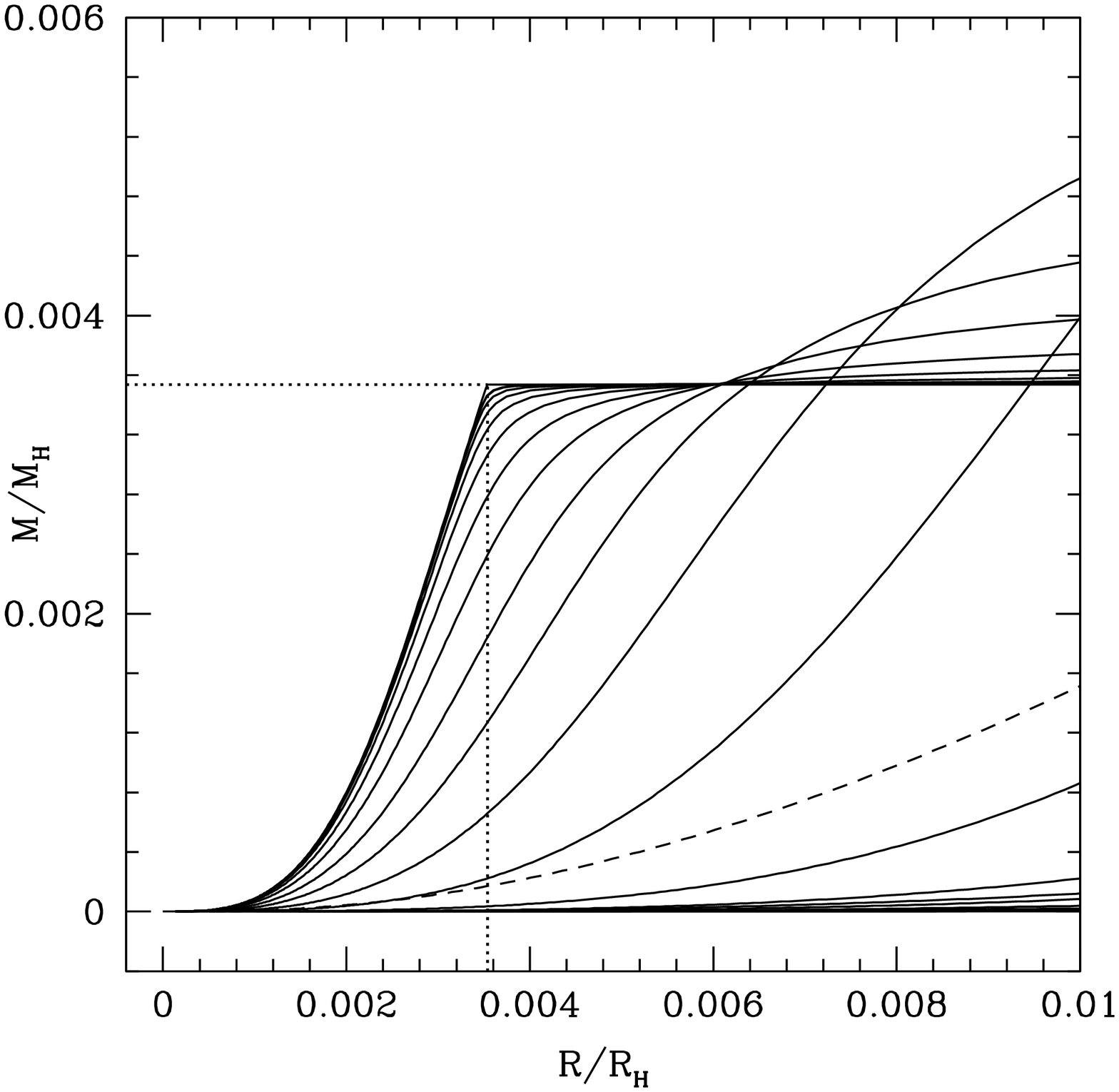}
 \caption{\label{fig.4}\small This figure shows further details for 
the same case as in figure \ref{fig.3}. The top two frames show the 
profiles of radial four velocity $U$ and energy density $e$ at two key 
moments: the time when the wind away from the central region reaches 
its maximum strength (solid line) and the moment when the void starts 
to refill from the outside (dot-dashed line). Note that the black hole 
is at a very small scale on the left-hand side of these plots. The 
bottom frame shows the mass profiles in the very central regions at 
different time levels, with the dashed curve representing the initial 
conditions used by the observer-time code and the dotted lines marking 
the mass and radius of the eventual black hole.}
 \end{figure}
%%%%%%%%%%%%%%%%%%%%%%%%%%%%%%%%%%%%%%%%%%%%%%%%%%

As material is shed from the central condensation when it is in the 
intermediate state, this forms a strong relativistic wind (as mentioned 
earlier) which excavates a deep semi-void around it. This is shown in the 
top two frames of figure \ref{fig.4} where the profiles of radial four 
velocity and energy-density are plotted. Two times are shown. The first 
(marked by the solid curves) is when the wind reaches its maximum 
strength, here with the maximum of $U/c$ being around 50 (for more extreme 
cases, we see values of more than 100). By this stage, a substantial 
semi-void has already formed around the central condensation and this then 
proceeds to deepen further as more material is blown outwards by the wind 
which is being accelerated by the steep pressure gradient at its inner 
edge. The second time shown (marked with the dot-dashed curves) is when 
the outward velocity has dropped to zero at the outer edge of the void and 
it then starts to refill. At this stage, the AMR scheme used for our 
simulations has reached 23 levels of refinement (a factor of almost 
$10^7$). When we follow the subsequent refilling of these voids, we find 
that it proceeds very gently and takes place on a timescale which is long 
compared to the dynamical timescale of the final collapse producing the 
black hole but short in cosmological terms. We then expect that further 
accretion onto the central black hole would proceed in a standard way. The 
bottom frame of figure \ref{fig.4} shows the profiles for the mass at 
successive times, with the dashed curve again marking the initial data for 
the observer-time code. Initially, when all of the medium is expanding, 
$M(R)$ is, of course, decreasing at all values of $R$ but later, when the 
collapse starts, $M(R)$ increases for the values of $R$ inside the 
collapsing region, giving asymptotically the final mass of the black hole, 
evaluated where $(2M/R)_{peak}$ tends to $1$ and the lapse $f$ tends to 0 
(this location is marked by the dotted lines). The effect of the void can 
be seen in the very flat profile of $M(R)$ outside this point (the last 
curve shown corresponds to the same time as the dot-dashed curves in the 
top two frames).

%%%%%%%%%%%%%%%%%%%%%%%%PICTURE%%%%%%%%%%%%%%%%%%%%%
\begin{figure}[ht!]
\centering
\includegraphics[width=6cm]{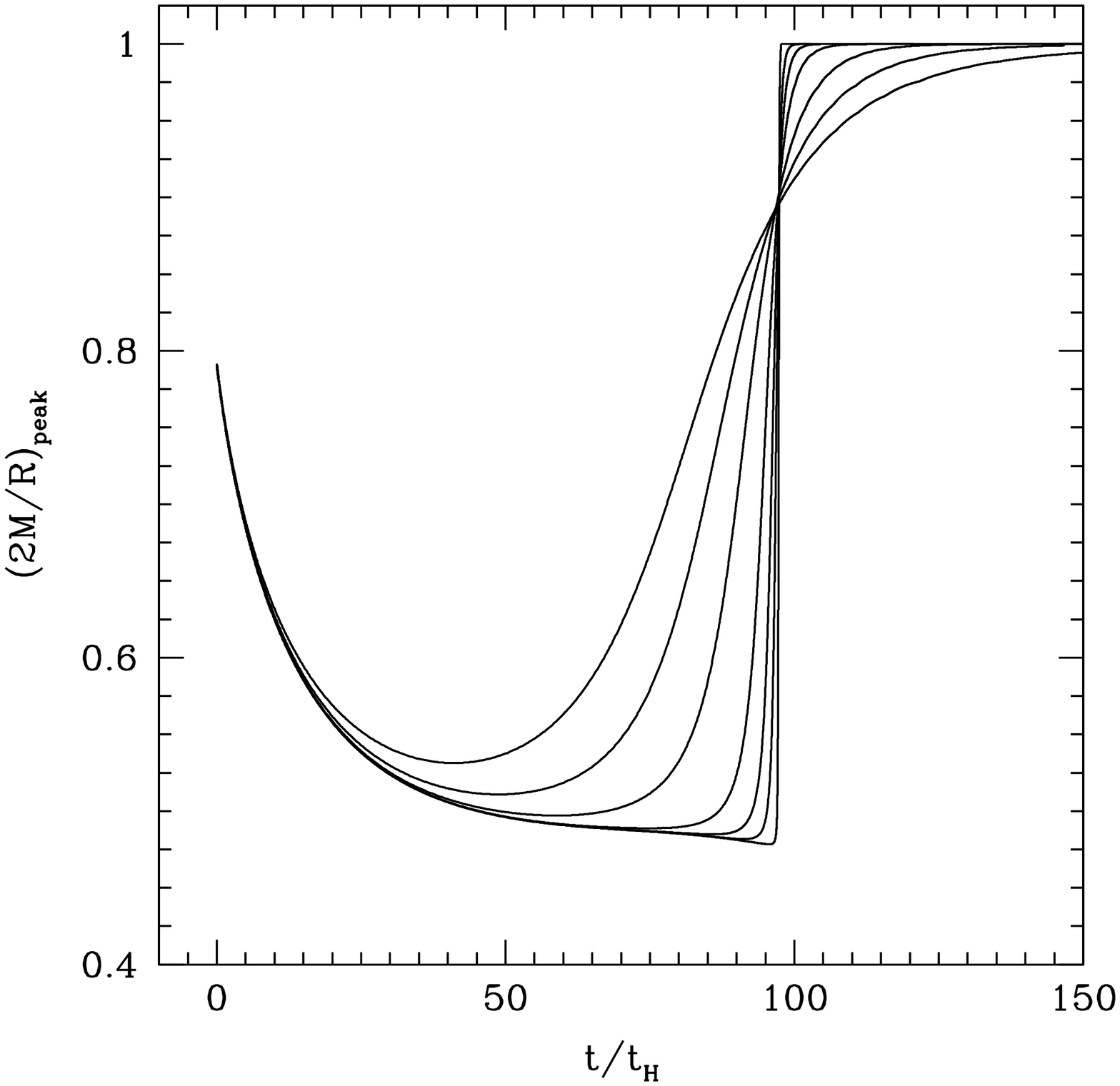}
\includegraphics[width=6cm]{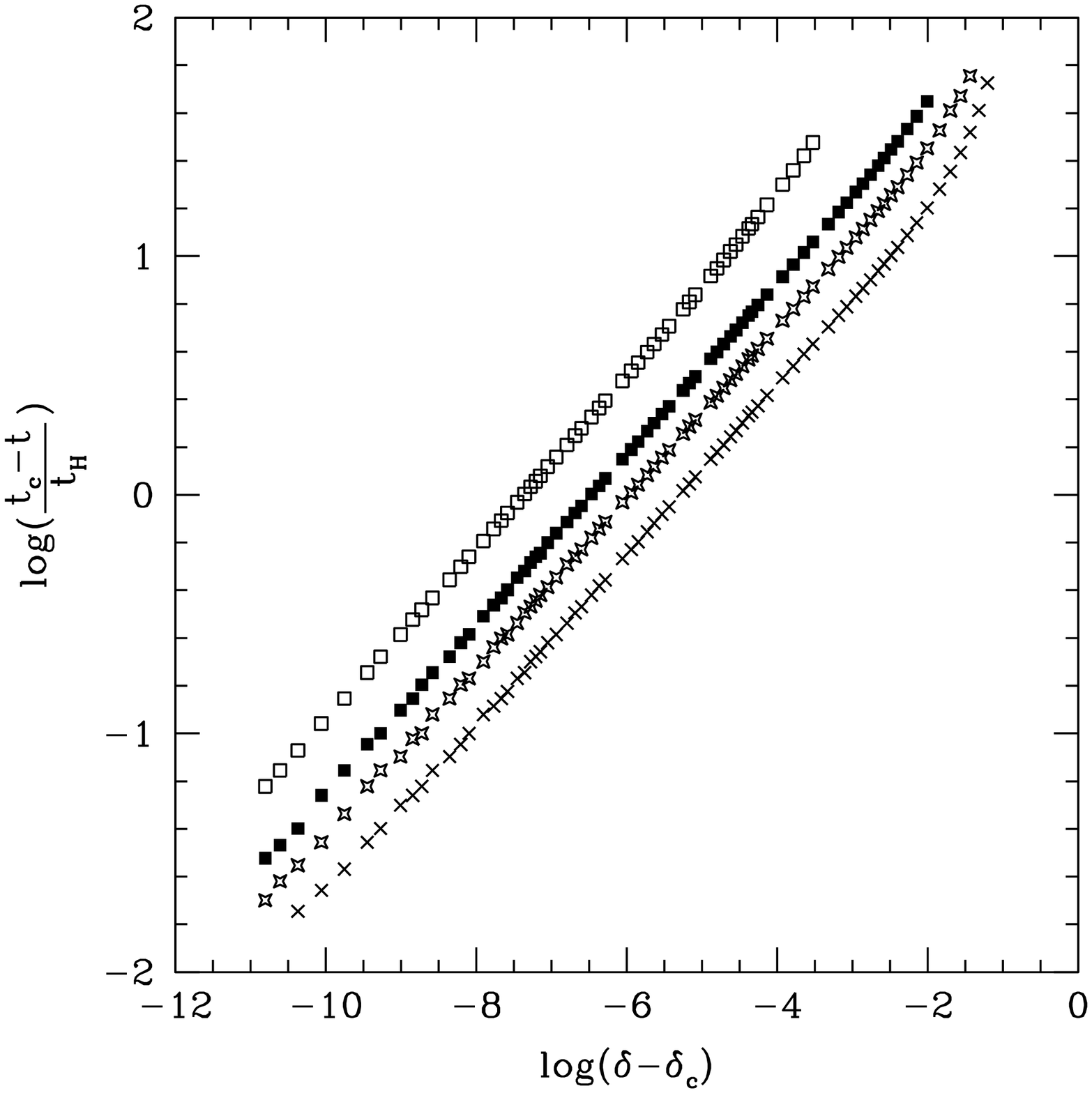}
 \caption{\label{fig.5}\small The left-hand frame shows 
$(2M/R)_{peak}$ plotted against the time $t$ (given in units of the 
horizon-crossing time $t_H$) for seven different cases of collapse 
forming a black hole. The right-hand frame shows a second scaling 
behaviour for the same cases as in figure \ref{fig.1}: plotting 
$(t_c-t)/t_H$ as function of $(\delta-\delta_c)$. The time $t$ used 
here is measured when $(2M/R)_{peak}$ becomes larger than 0.5, 0.6, 
0.7 and 0.8 respectively (the larger values corresponding to lower 
sets of points). Each of the four sets of data is fitted by a scaling 
law with $\gamma \simeq 0.36$.}
 \end{figure}
%%%%%%%%%%%%%%%%%%%%%%%%%%%%%%%%%%%%%%%%%%%%%%%%%%

Considering now the range of collapse models, it is interesting to plot 
$(2M/R)_{peak}$ as function of time and we do this, in the left-hand frame 
of figure \ref{fig.5}, for seven representative cases from the set shown 
in figure \ref{fig.1}. (The time $t$ is given here in units of the 
horizon-crossing time $t_H$. We recall that $t$ is the time as measured by 
a comoving observer at the outer edge of our grid). At the early times, 
just after the start of the observer-time calculation, the values of 
$(2M/R)_{peak}$ are almost identical for the different cases due to the 
rather small range of values of $\delta$, but differences grow as the 
evolution proceeds. (The values of $(\delta - \delta_c)$ decrease going 
from top to bottom of the curves on the left-hand side of the plot.) The 
time spent in the intermediate state at almost constant $(2M/R)_{peak}$ 
increases as $\delta$ tends to $\delta_c$ but reaches a finite limit. When 
gravity wins against pressure, causing departure from the intermediate 
state, the increase of $(2M/R)_{peak}$ towards 1 occurs with progressively 
steeper gradients as $\delta$ tends to $\delta_c$, becoming vertical in 
the limit. One can see that the curves all intersect at a particular point 
having $(2M/R)_{peak} \simeq 0.9$ and whose time coordinate is that of the 
vertical rise seen in the limit. We refer to this as the critical time 
$t_c$ and its value here is given by $(t_c/t_H \simeq 97.38)$. It is the 
time that would be taken in the critical limit $(\delta \to \delta_c)$ for 
all of the matter to be shed from the central condensation.

As noted previously, when using observer time, a black hole is fully 
formed only asymptotically as $t \to \infty$ and so one cannot speak of a 
finite time for black-hole formation. However, there is no problem in 
measuring the time at which $(2M/R)_{peak}$ rises through any particular 
threshold value between 0.5 and slightly less than 1 during the final 
approach to the black-hole state. As can be seen from the left-hand frame 
of figure \ref{fig.5}, this time gets progressively closer to $t_c$ as 
$\delta \to \delta_c$. Defining $t$ now to be the time at which a 
particular threshold is crossed, one finds that the values of $(t_c-t)$ 
satisfy scaling laws in $(\delta-\delta_c)$ for $\delta$ sufficiently 
close to $\delta_c$, as can be seen in the right-hand frame of figure 
\ref{fig.5} where we have plotted data for all of the cases shown in 
figure \ref{fig.1}. The four sets of points are for thresholds of 
$(2M/R)_{peak}$ equal to 0.5, 0.6, 0.7 and 0.8 respectively, going from 
top to bottom. Each of the scaling laws has $\gamma \simeq 0.36$, in close 
agreement with that for the black hole masses. This is consistent with 
what one would expect if this process is genuinely an example of the 
critical collapse phenomenon.

%%%%%%%%%%%%%%%%%%%%%%%%PICTURE%%%%%%%%%%%%%%%%%%%%%
\begin{figure}[ht!]
\centering
\includegraphics[width=6cm]{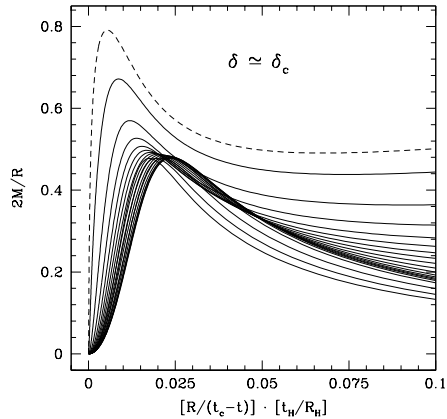}
 \caption{\label{fig.6} \small The $2M/R$ evolution for the critical 
case, shown in figure \ref{fig.2}, is re-plotted using 
$[R/(t_c-t)]\cdot[t_H/R_H]$ on the x-axis so as to look for 
self-similar behaviour.}
 \end{figure}
%%%%%%%%%%%%%%%%%%%%%%%%%%%%%%%%%%%%%%%%%%%%%%%%%%
  
Another characteristic feature of critical collapse is that it is related 
to a self-similar behaviour (see the review in \cite{Gundlach}). This 
means that it should be possible to write the solution in terms of 
suitable variables in such a way that it remains the same, independent of 
time, sufficiently close to the critical limit. In figure \ref{fig.6}, we 
plot $2M/R$ for the critical case, as shown in figure \ref{fig.2}, against 
$R/(t_c-t)$. Once again, the dashed curve represents the initial time. 
When the solution reaches the intermediate state, the inner parts do get 
progressively closer to reaching an unchanging profile as time goes on. 
The self similarity is not exact, but this is understandable since the 
collapse is occurring within an outer expanding medium with which there is 
a continuing interaction because of the shedding of matter from the 
central condensation. We have only a local approximate self-similar 
behaviour, but this is enough for having an associated scaling-law 
behaviour.

%===============================================================

%===============================================================

\section{Conclusions}
 In this paper, we have presented results from our investigation of the 
extent to which primordial black hole formation in the radiative era of 
the early universe can be considered as a manifestation of the critical 
collapse phenomenon. For doing this we have made numerical simulations 
including both the collapse producing the black hole and the continuing 
expansion of the surrounding universe, focusing on the behaviour of 
initial perturbations of a type which could have come from inflation, 
having only a growing component and no decaying component. Implementing an 
AMR scheme within our Lagrangian code, we have been able to follow 
formation of black holes over a range of more than a thousand in mass. For 
the type of perturbation that we are studying, we find that scaling-law 
behaviour persists down to the smallest masses that we are able to follow, 
with no sign of a levelling-off such as had been reported in earlier work. 
We attribute this difference to the different type of initial conditions 
used; the present ones do not lead to the formation of shocks during 
collapses giving rise to black holes, which were a key feature of previous 
work.

For cases near to the critical limit, we observe formation of a rather 
long-lived intermediate state with a central condensation having 
compactness $2M/R$ about half that of a black hole. This central 
condensation progressively sheds material in the form of an 
ultra-relativistic wind, excavating a semi-void between it and the 
surrounding matter of the universe, until it eventually either collapses 
to form a black hole or disperses into the surrounding medium. In cases 
where a black hole does form, we find that the void eventually refills 
rather gently, and we expect that subsequent accretion would then proceed 
in a standard way.

In addition to the scaling law for the black hole mass as a function of 
the closeness of the perturbation amplitude to the critical limit, we have 
also observed other features characteristic of critical collapse: 
existence of a critical surface, another associated scaling law with the 
same exponent, and an approximate similarity solution. Taken together, all 
of this leads us to conclude that we genuinely have here a potential 
physical realization of the critical collapse phenomenon. For 
perturbations of the type coming from inflation, we have not seen any 
evidence to make us think that there would be a deviation away from the 
scaling law at lower masses until either there is a change in the matter 
model or the quantum regime is reached.

\ack We gratefully acknowledge helpful discussions with a number of 
colleagues during the course of this work, including Bernard Carr, Carsten 
Gundlach, Ian Hawke and Karsten Jedamzik.

%%%%%%%%%%%%%%%%%%%%%%%%%%%%%%%%%%%%%%%%%%%%%%%%%%%%%%%%%%%%%%%
\
%%%%%%%%%%%%%%%%%%%%%%%%%%%%%%%%%%%%%%%%%%%%%%%%%%%%%%%%%%%%%%%%% 

\appendix
\section*{Appendix A: Corrections}
\setcounter{section}{1}

We have noted typographical errors in some of the equations in Paper 
2. We list the corrected forms below. The formulae used in the 
numerical calculations presented in Paper 2 were correct and there is 
no change in the results.

\vspace{12pt}\noindent Eq.(95):
\beq
K^\pr(r) = \frac{r}{\Delta^2}\lP \alpha - \lp
1+\alpha \frac{r^2}{2\Delta^2}\rp \rP \exp{\lp
-\frac{r^2}{2\Delta^2}\rp}
\eeq

\vspace{12pt}\noindent Eq.(100):
\beq
\frac{F(\alpha)}{2\alpha} \ug 3 \quad \mr{if} \quad \alpha \ug 0
\eeq

\vspace{12pt}\noindent Eq.(101):
\bea
&\!\!\!\!\!\!\!\!\!\!\!\!\!\!\!\!\!\!\!\!\!\!\!\!\!\!\!\!\!\!\!\!\!
\tilde{e}(Z)  \ug  \Phi(\xi) \frac{\Delta^2}{2\alpha}
F(\alpha) \lP 1 \,+\, \lp\frac{5}{6}\alpha-\frac{1}{3}\rp
\frac{F(\alpha)}{2\alpha}Z^2 \,-\, \frac{\alpha}{6}
\lp\frac{F(\alpha)}{2\alpha}\rp^{\!2} Z^4 \rP
\exp{\lp -\frac{F(\alpha)}{4\alpha}Z^2\rp}\nonumber\\
&
\eea

\vspace{12pt}\noindent Eq.(103):
\beq
\tilde{e}(Z)  \ug  2\Delta^2 \lP 1 - Z^2 \rP
\exp{\lp -\frac{3}{2}Z^2\rp}
\eeq

%%%%%%%%%%%%%%%%%%%%%%%%%%%%%%%%%%%%%%%%%%%%%%%%%%%%%%%%%%%%%%%
\
%%%%%%%%%%%%%%%%%%%%%%%%%%%%%%%%%%%%%%%%%%%%%%%%%%%%%%%%%%%%%%%%% 

\appendix
\section*{Appendix B: Convergence test of the code}
\setcounter{section}{2}

As noted in Section 3, our code has a long history and has been carefully 
tested in its various forms. Its approach and basic structure are very 
similar to those of the code for which extensive test-bed calculations 
were presented by Baumgarte, Shapiro \& Teukolsky \cite{Baumgarte} and we 
mentioned a number of other tests ourselves in Paper 2. In this Appendix, 
we describe a convergence test which we have carried out with our updated 
version of the code. In our view, more stringent tests are provided by 
carefully considering internal consistency features of the results, which 
we have done over a range of more than ten orders of magnitude in ($\delta 
- \delta_c$), but having a convergence test is nevertheless useful in 
telling more about the behaviour of the code and in estimating error bars.

Performing a convergence test presents some unusual aspects with the kind 
of code being used here, which has a non-uniform Lagrangian grid as its 
basic one, rather than an Eulerian grid, and which then has AMR on top of 
this. However, it is straightforward to proceed by dividing each zone of 
our standard grid into two equal parts and halving the refinement criteria 
on $\Delta R/R$, mentioned in Section 3, so as to produce a ``double 
resolution'' version of the code with which the calculation can then 
proceed as before. Similarly, a further subdivision can be made so as to 
produce a ``quadruple resolution'' version. Making a convergence test by 
comparing calculations done in this way is somewhat imperfect but it has 
the merit of being a test of the full code as it actually runs.

A particular point requiring care is in ensuring that the models being 
compared at different resolutions are genuinely equivalent. As discussed 
in Section 4, the hydrodynamic behaviour depends sensitively on the value 
of $(\delta - \delta_c)$, and we are going down as low as $10^{-11}$ in 
this quantity. When calculating the threshold $\delta_c$ with our three 
different resolutions, one finds values that vary in the fourth 
significant figure (although they are converging rapidly with increasing 
resolution - by about one extra significant figure with each halving of 
the grid). Since it is the difference between the actual $\delta$ and the 
threshold $\delta_c$ which determines the hydrodynamical behaviour, it is 
important that the models being compared at the different resolutions 
should all have the same $(\delta - \delta_c)$, but with the $\delta_c$ 
being subtracted from the $\delta$ always being calculated at the same 
resolution as the $\delta$. For seeing the convergent behaviour, it is 
necessary to have a sufficient number of reliable significant figures in 
the results. The higher resolutions give very precise results but do not 
allow calculations to be made for values of $(\delta - \delta_c)$ as small 
as those possible with the standard resolution, mainly because of running 
into cancellation errors in the calculation of the pressure gradients. 
With the quadruple resolution, we can determine $\delta_c$ correct to nine 
significant figures but no more. This, then, determines the accuracy to 
which we can say that the models being compared at the different 
resolutions have the same $(\delta - \delta_c)$. In order that the models 
being compared have the same $(\delta - \delta_c)$ correct to five 
significant figures (as necessary for making the convergence test), we 
cannot then use a case with $(\delta - \delta_c) < 10^{-4}$.

%%%%%%%%%%%%%%%%%%%%%%%%PICTURE%%%%%%%%%%%%%%%%%%%%%
\begin{figure}[ht!]
\centering
\includegraphics[width=6.1cm]{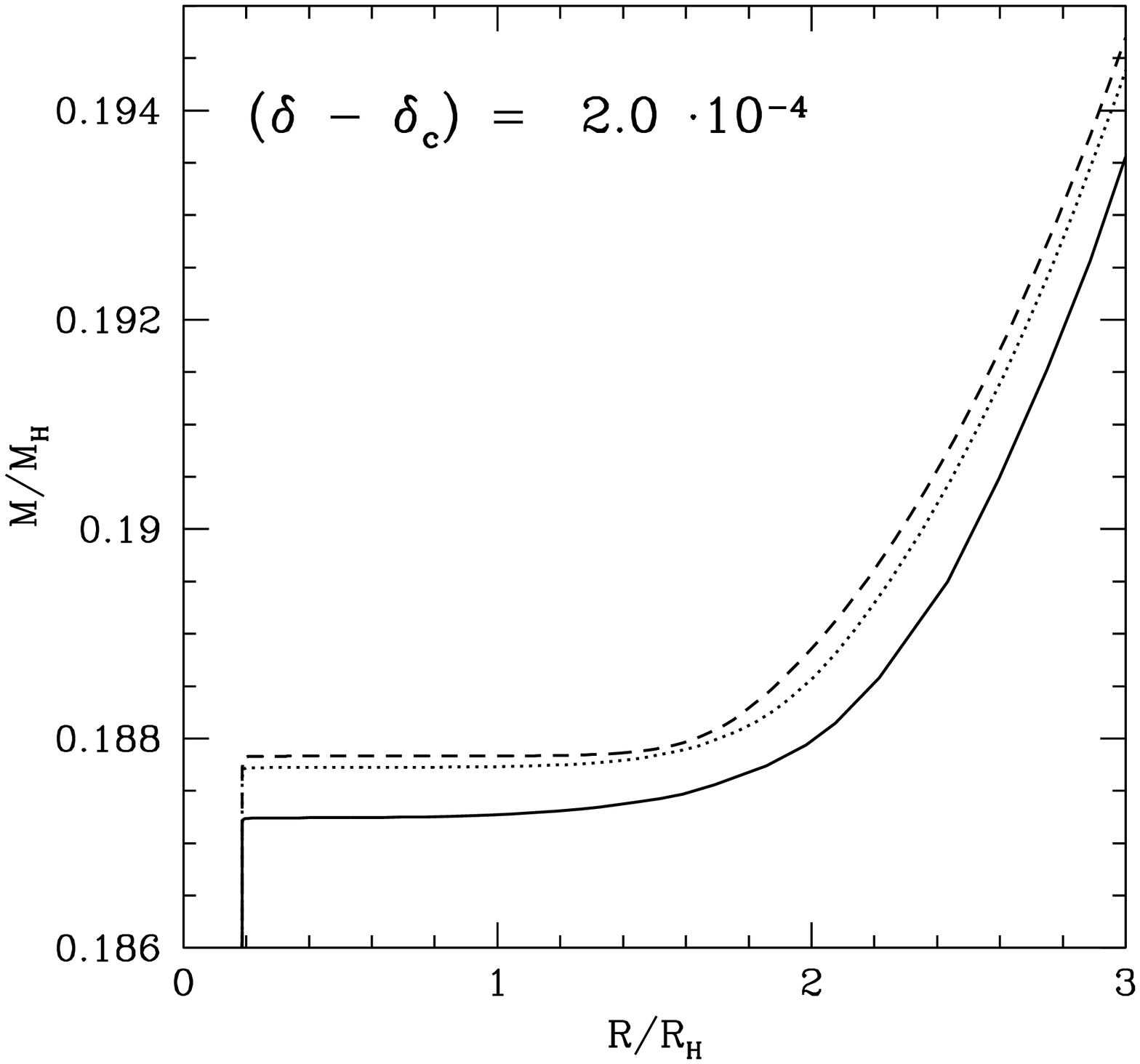}
 \caption{\label{fig.B1} \small Convergence test for the code: profiles 
are shown of $M$ against $R$ for $(\delta - \delta_c) = 2\cdot10^{-4}$, as 
calculated with the standard resolution used for the calculations 
presented in this paper (solid line), double resolution (dotted line) and 
quadruple resolution (dashed line). Note the very expanded vertical 
scale.}
 \end{figure}
%%%%%%%%%%%%%%%%%%%%%%%%%%%%%%%%%%%%%%%%%%%%%%%%%%

In figure \ref{fig.B1}, we show the results of a convergence test made 
with $(\delta - \delta_c) = 2\cdot10^{-4}$, plotting $M$ against $R$ at 
the three resolutions, with the standard resolution being shown with the 
solid line, the double resolution with the dotted line, and the quadruple 
resolution with the dashed line. Note that the vertical scale in this plot 
has been greatly expanded so that the differences can be easily seen. The 
curves are clearly converging with an order of convergence of around two 
although it is not quite uniform, as expected with the way in which the 
test is being done. The resulting error bars are at a fraction of one 
percent, so that the difference between results with the different 
resolutions would just barely be visible on the other plots presented in 
this paper. The main noticeable effect is that the curves become slightly 
smoother.

The case used for the test is not one of the most extreme ones (for the 
reason given above) but the AMR is still going through eleven levels of 
refinement and so the test is a serious one. The region shown in figure 
\ref{fig.B1} is the one where the differences with resolution can be most 
clearly seen. The behaviour of the scaling law is the same at all of the 
resolutions.

%%%%%%%%%%%%%%%%%%%%%%%%%%%%%%%%%%%%%%%%%%%%%%%%%%%%%%%%%%%%%%%
\
%%%%%%%%%%%%%%%%%%%%%%%%%%%%%%%%%%%%%%%%%%%%%%%%%%%%%%%%%%%%%%%%% 

\end{document}